\documentclass[12pt]{article}

\usepackage{amsmath, amssymb, slashed, epsf, color, graphicx}

\usepackage{subfigure}

\begin{document}

\begin{titlepage}
\begin{center}

\hfill IPMU-13-0084 \\
\hfill ICRR-Report-652-2013-1\\
\hfill \today

\vspace{1.5cm}
{\large\bf
AMS-02 Positrons from Decaying Wino \\
in the Pure Gravity Mediation Model\\
}

\vspace{2.0cm}
{\bf Masahiro Ibe}$^{(a, b)}$,
{\bf Shigeki Matsumoto}$^{(b)}$, \\
{\bf Satoshi Shirai}$^{(c)}$
and
{\bf Tsutomu T. Yanagida}$^{(b)}$ \\

\vspace{1.5cm}
{\it
$^{(a)}${Institute for Cosmic Ray Research (ICRR), Theory Group, \\
University of Tokyo, Kashiwa, Chiba 277-8568, Japan} \\
$^{(b)}${Kavli Institute for the Physics and Mathematics of the Universe (IPMU),\\
University of Tokyo, Kashiwa, Chiba 277-8568, Japan} \\
$^{(c)}${Berkeley Center for Theoretical Physics, Department of Physics, \\
and Theoretical Physics Group, Lawrence Berkeley National Laboratory, \\
University of California, Berkeley, CA 94720, USA} \\
}

\vspace{1.5cm}
\abstract{
The AMS-02 collaboration has recently reported an excess of the cosmic-ray positron fraction, which turned out to be consistent with previous results reported by the PAMELA and Fermi-LAT collaborations. A decaying dark matter with the mass around 1~TeV can be responsible for the excess of the positron fraction when it is interpreted as a dark matter signal. Interestingly, the pure gravity mediation model provides such a dark matter, namely an almost pure neutral wino dark matter, when a tiny R-parity violation through $LLE^c$ interactions is introduced. We show that the decaying wino dark matter well reproduces the energy spectrum of the fraction with being consistent with constraints from cosmic-ray anti-proton and gamma-ray observations.
}

\end{center}
\end{titlepage}
\setcounter{footnote}{0}

\section{Introduction}
\label{sec: introduction}

The AMS-02 collaboration recently released their first result of the cosmic-ray positron fraction~\cite{AMS-02}. They reconfirmed the anomalous excess of the positron fraction over the expectation based on simple cosmic-ray propagation models, which had been found by the PAMELA~\cite{Adriani:2008zr} and Fermi-LAT~\cite{FermiLAT:2011ab} collaborations. One interesting fact the AMS-02 collaboration found is that the positron fraction shows no observable anisotropy. Though it is obviously premature to make a definite statement on it because of limited statistics, this fact encourages several dark matter interpretations of the excess rather than astrophysical interpretations.

When the excess of the positron fraction is interpreted as a dark matter signal, a decaying dark matter with the mass around 1~TeV could be responsible for the excess. In particular, the decay of a wino-like neutralino dark matter is known to naturally account for such an excess as discussed in reference~\cite{Shirai:2009fq}. The wino-like dark matter is a natural prediction when the gaugino masses are dominated by the anomaly mediated supersymmetry (SUSY) breaking contributions~\cite{AMSB}. Such scenarios have been widely discussed in the models of pure gravity mediation~\cite{PGMs, PGM2}, the models with strong moduli stabilization~\cite{Strong Moduli}, the spread supersymmetry~\cite{spread}, and the minimal split supersymmetry~\cite{Split SUSY}, all of which explain the observed mass of about 126~GeV for the Higgs boson~\cite{OYY}. Thanks to the relatively large gravitino mass $m_{3/2}= {\cal O}$(100--1000)~TeV in these models, almost all of the phenomenological and cosmological problems in SUSY standard model are solved. The mass of the neutral wino dark matter is predicted to be $m_{\rm wino} = {\cal O}$(0.1--1)~TeV. In particular, the wino with the mass $m_{\rm wino} \simeq 2.7$~TeV is interesting since its thermal relic can explain the observed dark matter density if it is stable or long-lived~\cite{Hisano:2006nn}.

In this paper, we show that the decaying wino dark matter with its mass around 1~TeV can naturally account for the anomalous excess of the positron fraction reported by the AMS-02 collaboration. In particular, if the decay occurs dominantly through R-parity violating interactions $LLE^c$, the energy spectrum of the positron fraction is well reproduced with being consistent with constraints from cosmic-ray anti-proton and gamma-ray observations. After briefly reviewing the wino dark matter in the framework of the pure gravity mediation model and discussing its decay processes through the $LLE^c$ interactions in section \ref{sec: wino dark matter}, we explicitly show the parameter region (the mass and the lifetime of the wino dark matter) which accounts for the AMS-02 result in section \ref{sec: decaying wino}. Section \ref{sec: summary} is devoted to summary and discussions of some prospects for testing the decaying wino dark matter in (near) future. In the appendix A, we also discuss a ``GUT" model for the R-parity violation.


\section{Wino dark matter in the PGM model}
\label{sec: wino dark matter}

We first summarize the wino dark matter predicted by the pure gravity mediation (PGM) model~\cite{PGMs} and its decay into electrons/positrons through $LLE^c$ interactions, where $L$ and $E^c$ are the lepton doublet and charged lepton singlet, respectively.\footnote{The analysis developed in the paper is also applicable to the most of models with heavy sfermion as long as the lightest supersymmetric particle (LSP) is the almost pure wino-like neutralino.} In the PGM model, all gauginos acquire their masses from one-loop anomaly mediated contributions~\cite{AMSB}, and the neutral wino becomes the LSP. The Higgsino mass term, i.e. $\mu$-term, of order of the gravitino mass is generated through interactions to the R-symmetry breaking sector~\cite{Inoue:1991rk}.\footnote{This effect is not exactly the same as the original Giudice-Masiero mechanism~\cite{Giudice:1988yz}, because the Higgsino mass term (the $\mu$-term) is not generated by interactions to the SUSY breaking sector.} Thus, with such a large $\mu$-term, the mixing between the wino and the bino is highly suppressed. The PGM model therefore predicts the almost pure neutral wino as a dark matter candidate.

The lower limit on the wino mass is currently set by the Large Hadron Collider experiment~\cite{ATLAS:2012mna} and the Fermi-LAT~\cite{Ackermann:2011wa} observation. In the former case, the direct charged wino production with a disappearing charged track~\cite{PGMs, Charged Track} gives a limit as $m_{\rm wino} \gtrsim 110$~GeV~\cite{Ibe:2012sx}. In the latter case, the observation of gamma-rays from wino dark matter annihilations at milky-way satellites (dwarf spheroidals) gives a limit as 2.2~TeV $\gtrsim m_{\rm wino} \gtrsim$ 500~GeV and $m_{\rm wino} \gtrsim$ 2.4~TeV~\cite{Sommerfeld}. On the other hand, the upper limit on the wino mass is obtained by cosmology. The wino dark matter is produced thermally~\cite{Hisano:2006nn} and non-thermally through gravitino decays\footnote{The non-thermal contribution depends also on the reheating temperature ($T_R$) after inflation, where it is estimated to be $\Omega^{({\rm NT})}_{\rm wino} h^2 \simeq 0.16 \times (m_{\rm wino}/300~{\rm GeV}) \times (T_R/10^{10}~{\rm GeV})$~\cite{Moroi:1999zb, PGMs}.} in the early universe, and both contribute to the dark matter density observed today. It then turns out that the wino mass must be lighter than $\sim$2.7~TeV in order not to be over-produced. The wino mass of around 2.7~TeV is particularly interesting because the dark matter density is explained only by its thermal relic~\cite{Hisano:2006nn}.

The wino dark matter decays into standard model particles when the R-parity is violated in the PGM model. There are several ways to introduce the R-parity violation, which makes the wino meta-stable. The simplest way is the use of $L_iH_u$ interactions with $H_u$ being the up-type Higgs doublet. Then the wino decays into a $Z$ boson plus a neutrino, a Higgs boson plus a neutrino, and a $W$ boson plus a charged lepton~\cite{Gravitino decay}. Though the decay processes indeed produce high-energy electrons/positrons, they also produce anti-protons because of cascade decays of $Z$, $W$ and Higgs bosons, leading to a severe limit on its lifetime~\cite{Garny:2012vt}. Furthermore, those cascade decays produce gamma-rays with a hard spectrum. High-energy charged leptons produced by the two-body decay processes also generate energetic gamma-rays through the inverse Compton scattering (ICS) with the cosmic microwave backgrounds and infra-photons from star lights~\cite{Ishiwata:2009dk}. Those facts make the decaying wino dark matter interpretation of the positron excess difficult~\cite{Cirelli:2012ut}.

\begin{figure}[t]
\begin{center}
\includegraphics[width=50mm]{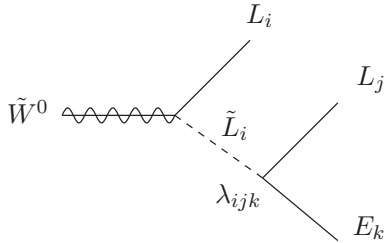}
\caption{\small \sl The decay of the wino dark matter through the interaction $\lambda_{ijk}L_iL_jE^c_k$.}
\label{fig: decay}
\end{center}
\end{figure}

In this letter, we therefore focus on the wino decay caused by the interactions,
\begin{eqnarray}
\label{eq:LLE}
{\cal W}_{\slashed{R}} = \lambda_{ijk} \, L_iL_jE^c_k,
\end{eqnarray}
where we do not have to worry about the constraint from cosmic-ray anti-proton observations, while another constraint from gamma-ray observations also becomes milder than the case of $LH_u$~\cite{Shirai:2009fq}.
\footnote{
There also be some studies discussing contributions to anti-proton and gamma-ray fluxes from next leading order processes such as the electroweak bremsstrahlung, where these contributions become important in some models aiming to explain the AMS-02 result \cite{DeSimone:2013fia}. In our model, with the wino mass of ${\cal O}(1)$ TeV, these processes are less significant~\cite{Ciafaloni:2010ti}.
}
The assumption that R-parity violation is dominated by $LLE^c$ interactions needs to be justified in the grand unified theory (GUT) in which quarks and leptons are unified into GUT multiplets. We discuss a model addressing the origins of the interactions in appendix \ref{sec: LLE interactions}, where only $LLE^c$ are generated while suppressing hadronic R-parity violating interactions in a GUT consistent way.

Through the R-parity violating interactions in Eq.\,(\ref{eq:LLE}), the wino dark matter decays into two charged leptons by emitting a neutrino through three-body processes as shown in figure~\ref{fig: decay}. The wino lifetime is then estimated to be~\cite{Barbier:2004ez}
\begin{equation}
\tau_{\rm wino} \sim
10^{27} \, [{\rm sec.}] \,
(\lambda/10^{-19})^{-2} \,
(m_{\rm wino}/1~{\rm TeV})^{-5} \,
(m_{\tilde L}/10^3~{\rm TeV})^4.
\label{lifetime}
\end{equation}
The decay can be a source of high-energy cosmic-rays which results in the excess of the positron fraction. In the next section, we discuss positron, electron and gamma-ray fluxes which are expected from the decaying wino dark matter.

\section{Signals of the decaying wino dark matter}
\label{sec: decaying wino}

For dark matter signals, we assume that the left-handed slepton of the third generation ($\tilde{L}_3$) is lighter than those of other generations ($\tilde{L}_1, \tilde{L}_2$) and the wino dark matter decay occurs dominantly by exchanging $\tilde L_3$. We also assume no flavor violation on the coupling between $\tilde{L}_i$, $L_i$ and $\tilde{W}^0$ for simplicity. In following discussions, we consider the cases of $\lambda_{32i}$ ($i = 1, 2, 3$) as a demonstration. Branching fraction of the decay is then given by ${\rm BF}(\tilde W^0 \to \tau_L \nu_\mu e_{R i}) = {\rm BF}(\tilde W^0 \to \mu_L \nu_\tau e_{R i}) = 0.5$. Subsequent decays of heavier leptons lead to energetic positrons, electrons, neutrinos and photons. High-energy photons are produced also from final state radiations. We estimate the primary spectra of the particles from these prompt decay channels using the Pythia~8 code~\cite{Sjostrand:2007gs} with a slight modification by us to treat the decay of a polarized lepton.

\subsection{Electron and positron fluxes}

In the calculation of electron and positron fluxes from the wino decay, the NFW profile~\cite{Navarro:1996gj} is used for the dark matter mass density of our galaxy with the profile parameters being $\rho_\odot = 0.4~{\rm GeV/cm^3}$ (the local halo density), $r_c = 20~{\rm kpc}$ (the core radius), and $r_\odot = 8.5~{\rm kpc}$ (the distance between our solar system and the galactic center). In order to take account of the effect of electron and positron propagations inside our galaxy, we have solved the diffusion equations according to the method in reference~\cite{Delahaye:2007fr}. We have mainly used the so-called MED model in the reference, but other models M1 and M2 have also been used to estimate the uncertainty associated with the propagations. The M1 (M2) model maximizes (minimizes) the fluxes with the Boron to Carbon ratio being consistent with its observation.

For astrophysical backgrounds against the signals, we must be careful about their calculations because of large uncertainties. In order to deal with the uncertainties, we have adopted a similar method developed in reference~\cite{Cirelli:2008pk}. Using four parameters $A^\pm$ and $p^\pm$, the
interstellar background fluxes are parameterized as
\begin{equation}
\Phi^{e^\pm}_{\rm BG}(E) =
A^\pm \, E^{p^\pm} \, \Phi^{e^\pm}_{\rm ref}(E),
\end{equation}
where $\Phi^{e^\pm}_{\rm ref}(E)$ are reference background fluxes, and we have adopted conventional ones obtained by the GALPROP code~\cite{Strong:1998pw} with the electron injection index being $-2.66$, which gives the best fit to the observed electron flux at PAMELA~\cite{Adriani:2011xv}. In order to take account of the effect of the solar modulation, which is important for low energy electrons and positrons, we have used the force-field method~\cite{Gleeson:1968zza} in both signal and background calculations.

We have fitted above electron and positron fluxes to several experimental data. The data we have considered are the electron flux at PAMELA~\cite{Adriani:2011xv}, the combined flux of electron and positron at Fermi-LAT~\cite{Ackermann:2010ij}, and the positron fraction at AMS-02~\cite{AMS-02}. In the fitting to Fermi-LAT data, we have included the systematic uncertainty associated with the absolute energy scale $S_{\rm Fermi}$. As a result, the electron and positron fluxes are fitted to the data by varying the following nine parameters: the wino lifetime ($\tau_{\rm wino}$), the wino mass ($m_{\rm wino}$), the background parameters ($A^\pm$, $p^\pm$), the force-field potentials for electrons and positrons ($\phi^\pm$), and the energy scale ($S_{\rm Fermi}$). We have varied the parameters $A^\pm$, $p^\pm$, and $\phi^\pm$ within the ranges of $A^\pm \in [0, \infty]$, $p^\pm \in [-0.5, 0.5]$, and $\phi^\pm \in [0, 1]$ GV, respectively, while adopted the error of the scale $S_{\rm Fermi}$ to be $\delta S_{\rm Fermi} = \pm 0.1$. All the errors associated with $S_{\rm Fermi}$ and experimental data are assumed to be Gaussian and we symmetrize the asymmetric error, if necessary. The fitting has been performed in the energy range of $E > 5$~GeV for AMS-02 positron fraction, $E > 10$~GeV for PAMELA electron flux, and $E > 20$~GeV for Fermi-LAT electron and positron total flux.

\subsection{Gamma-ray fluxes}

We have considered several types of gamma-ray fluxes from the decaying wino dark matter in order to investigate how severely the mass and width of the wino dark matter are limited. The fluxes we have calculated are the following: The isotropic component of the diffuse gamma-ray flux, the diffused gamma-ray flux from all sky including the region of galactic center but not including the region of galactic plane, gamma-rays from galactic clusters. We have also considered limits on the mass from the wino annihilation, which are also obtained from several gamma-ray observations. In what follows, we explain how the gamma-ray fluxes have been calculated and how we have put limits on the mass and width of the wino.

\subsubsection{Isotropic component of the diffuse $\gamma$-ray}

In order to calculate the isotropic component of the diffuse gamma-ray flux from the wino decay, 
both galactic and extra-galactic contributions have to be considered. 
For the galactic contribution, we have calculated the gamma-ray flux in the direction of the anti-galactic center 
according to the dark matter profile mentioned above. 
To estimate the galactic ICS contribution, we need to know how the electrons and positrons are diffused globally in the galactic halo.
However, the electron/positrons fluxes observed by PAMELA/Fermi/AMS-02 are local ones and not necessarily relate to the global abundance.
Its estimation has ${O}(1)$ uncertainties, depending on the assumption of the diffusion models.
Therefore, we have not included the contribution from the ICS in the galactic halo, to be cautious of the large uncertainties.

For extra-galactic component, 
we have calculated the contribution from the wino decay at all past redshifts, where both prompt decay and ICS with the cosmic microwave background are considered. The calculated flux has been compared with the experimental data officially reported by the Fermi-LAT collaboration~\cite{Abdo:2010nz}, and we have put a limit on the mass and width of the wino by demanding that the signal flux from the wino decay does not exceed any Fermi-LAT data points by more than two sigma significance.

\subsubsection{All sky survey of the diffuse $\gamma$-ray}

For the all sky survey, the calculation of the diffused gamma-ray flux from the wino decay is essentially the same as that of the isotropic component. Only the difference is that the flux is now calculated at all directions. We have considered both galactic and extra-galactic contributions in the same manner mentioned above. The calculated flux has been compared with another experimental data reported again by the Fermi-LAT collaboration~\cite{Ackermann:2012qk}. We have considered almost of all region of the sky including the region of galactic center but excluding other regions along the galactic plane. The strategy to put a limit on the mass and width of the dark matter is also the same as in the case of isotropic diffused gamma-rays.

\subsubsection{$\gamma$-ray from galactic clusters}

We have focused on the Fornax cluster as a target region to detect the gamma-ray signal, because this cluster is known to give the most severe limit on decaying dark matters among several clusters~\cite{Dugger:2010ys, Huang:2011xr}. 
In the calculation of the signal flux, we have included the effect concerning the extent of the dark matter profile, which results in a weaker limit than that using a point-source approximation. We have assumed the NFW profile for the dark matter density inside the cluster and estimated the profile parameters $\rho_s$ and $r_s$ with use of Virial concentration parameter ($c_{\rm vir}$)~\cite{Buote:2006kx, Duffy:2008pz}, Virial over-density parameter ($\Delta_{\rm vir}$)~\cite{Hu:2002we}, and the cluster mass \cite{Reiprich:2001zv, Chen:2007sz} according to reference~\cite{Hu:2002we}. Though ${O}(1)$ uncertainty exists in the estimation, we have adopted the following values for the profile parameters: $\rho_s = 0.09~{\rm GeV/cm^3}$ and $r_s = 0.14~{\rm Mpc}$. In this setup, the so-called astrophysical $J$-factor for the signal flux with a radius of one degree centered at the target is $J \, d\Omega = 1.9 \, \times \, 10^{19} \, {\rm GeV \, cm^{-2}}$.

In order to estimate the background from the diffuse gamma-ray, we have used {\tt iso\_p7v6source.txt} for its isothermal component and {\tt gal\_2yearp7v6\_v0.fits} for its galactic component, where both are provided by the Fermi-LAT collaboration~\cite{Fermi: BG}. 
For the background from point sources, we have varied only the parameters of the source 
J0334.3-3728 very close to the Fornax cluster as nuisance parameters.
Parameters of other point sources in the 2FGL catalog have been fixed to the best fit values under background only hypothesis. With use of {\tt make2FGLxml.py}~\cite{Fermi: USER}, we have included 2FGL sources within a radius of fifteen degrees as the background.

We then put a limit through a signal plus background fit. For the experimental data, we have used {\tt P7SOURCE\_V6} event class of the Fermi-LAT data measured between August 4, 2008 and April 21, 2013 in the energy region $E >$ 400~MeV, and in the region of interest (ROI) with a radius of ten degrees centered at the cluster. We have selected events passing the conditions recommended by the Fermi-LAT collaboration, using the package {\tt Fermi Science Tools}~\cite{Fermi: TOOL}.

\subsubsection{$\gamma$-ray from the wino annihilation}

We have also considered the limit on the wino mass from the wino annihilation. This is because that the annihilation cross section of the wino (mainly into $W$ boson pair) is significantly enhanced in some ${\cal O}(1)$ TeV mass region. The most severe limit is again from several gamma-ray observations at Fermi-LAT. We have therefore included the limits in our analysis obtained from observations of dwarf spheroidal galaxies~\cite{Ackermann:2011wa, Ando:2012vu}, galactic clusters~\cite{Ackermann:2010rg, Ando:2012vu}, and diffuse photons~\cite{Abdo:2010nz, Ackermann:2012qk}.

\subsection{Results}

\begin{figure}[t]
\begin{center}
\includegraphics[width=150mm]{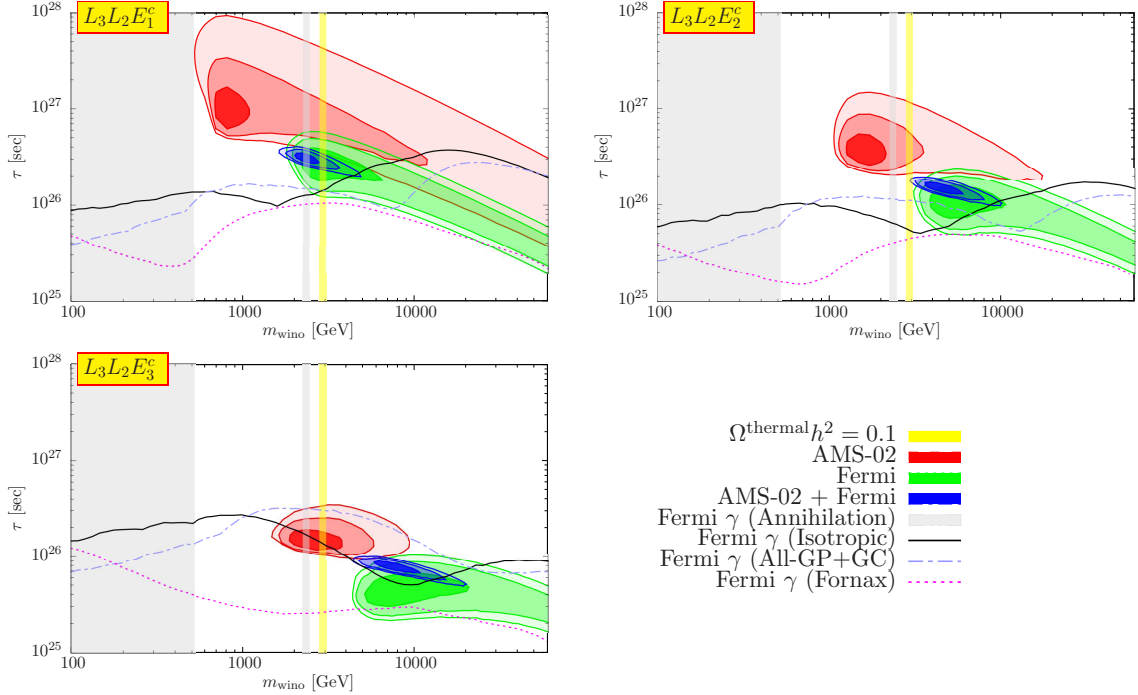}
\caption{\small \sl Fitting results and limits from $\gamma$-ray observations. See text for explanation.}
\label{fig: fit}
\end{center}
\end{figure}

\begin{figure}[p]
\begin{center}
\includegraphics[width=150mm]{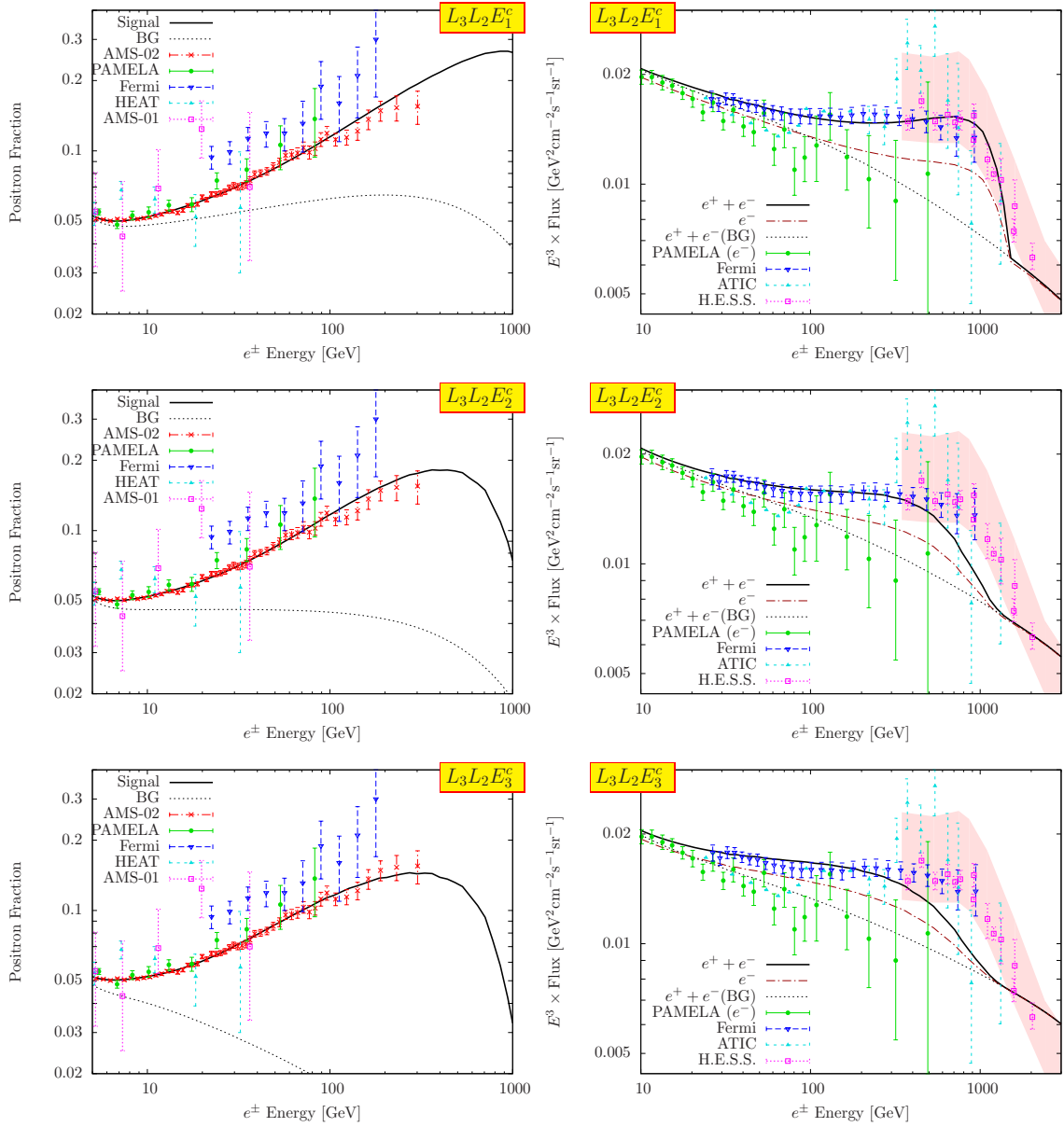}
\caption{\small \sl
Some examples of cosmic-ray signals with the wino mass of $m_{\rm DM} = 3$~TeV: Left panels show the predicted positron fraction and experimental data~\cite{AMS-02, Adriani:2008zr, FermiLAT:2011ab, Barwick:1997ig, Aguilar:2007yf}, while right panels show the electron and/or positron flux~\cite{Adriani:2011xv, Ackermann:2010ij, Collaboration:2008aaa, Aharonian:2009ah, :2008zz}. The shaded region shows the systematics error associated with the scale of absolute energy at the H.E.S.S. experiment. The best fitted Fermi energy scale is taken in the plot.}
\label{fig: sample}
\end{center}
\end{figure}

We are now at the position to discuss out results on the electron/positron flux and limits from gamma-ray observations, all of which have been obtained from numerical computations based on the strategy mentioned in the previous subsection. In figure~\ref{fig: fit}, the fitting result is depicted in each panel as contour lines of $\Delta \chi^2 =$ 2.3, 6.0, and $9.2$, corresponding to 68th, 95th, and 99th percentile of the Chi-Squared distribution with two degrees of freedom, respectively. The red, green, and blue contours are the results of fitting to the AMS-02, Fermi-LAT, and both the AMS-02 \& Fermi-LAT data. The PAMELA data of the electron flux is always used in the all fittings, where it plays an important role to fix the normalization of the background electron flux. The limits from gamma-ray observations at the Fermi-LAT experiment~\cite{Abdo:2010nz, Ackermann:2010qj} are also shown in each panel, where the black solid line named ``{\bf Fermi $\gamma$ (Isotropic)}" and the blue dash-dotted line named ``{\bf Fermi $\gamma$ (All-GP+GC)}" are from the isotropic component and all sky survey of the diffuse gamma-ray observations, while the pink dotted line named ``{\bf Fermi $\gamma$ (Fornax)}" is from the gamma-ray observation of the Fornax cluster. The limit form the wino annihilation is shown as a grey shaded region named ``{\bf Fermi $\gamma$ (Annihilation)}". The best fitted mass and width of the wino are summarized in table~\ref{tb: fit} with the absolute value of the $\chi^2$ divided by the degree of freedom. For convenience to estimate the uncertainty associated with electron and positron propagations in our galaxy, the fitting results using M1 and M2 diffusion models are also shown in the table. One can see that the uncertainty of the propagation model does not change the result drastically. In figure \ref{fig: sample}, we have also plotted the positron fraction and the combined electron plus positron flux for the case of best fit parameters with keeping the wino mass at 3~TeV.

\begin{table}[t]
\begin{center}
\caption{\small \sl Fit results in some selected parameter space. See text for explanation.}
\label{tb: fit}
\vspace{0.2cm}
\begin{tabular}{|c|c|c|c|c|}
\hline
Mode & Data & $m_{\rm wino}$ [GeV] & $\log_{10}(\tau~{\rm [sec]})$  & $\chi^2/dof$ \\
\hline
Background & AMS-02 & - &  -  & $ 55.1/67 $ \\
\hline
Background & Fermi & - &  -  & $ 47.8/43 $ \\
\hline
Background & AMS-02 + Fermi & - &  -  & $ 271.8/93 $ \\
\hline
\hline
$L_3 L_2 E^c_1$ & AMS-02 & 806 &  26.8  & $ 43.7/65 $ \\
\hline
$L_3 L_2 E^c_1$  & Fermi & 2611 &  26.3  & $ 26.3/41 $ \\
\hline
$L_3 L_2 E^c_1$  & AMS-02 + Fermi &  2290  & 26.4 & $ 65.9/91 $ \\
\hline
$L_3 L_2 E^c_1 $  & AMS-02 + Fermi &  3000 (fixed)  & 26.4 & $ 69.2/92 $ \\
\hline
\hline
$L_3 L_2 E^c_2$ & AMS-02 & 1581 &  26.5  & $ 40.3/65 $ \\
\hline
$L_3 L_2 E^c_2$  & Fermi & 5055 &  25.9  & $ 12.3/41 $ \\
\hline
$L_3 L_2 E^c_2$  & AMS-02 + Fermi & 4209 &  26.1  & $ 62.4/91 $ \\
\hline
$L_3 L_2 E^c_2 $  & AMS-02 + Fermi &  3000 (fixed)  & 26.1 & $ 70.3/92 $ \\
\hline
\hline
$L_3 L_2 E^c_3$ & AMS-02 & 2784 &  26.0  & $ 36.1/65 $ \\
\hline
$L_3 L_2 E^c_3$  & Fermi & 7439 &  25.5  & $ 25.5/41 $ \\
\hline
$L_3 L_2 E^c_3$  & AMS-02 + Fermi & 7957 &  25.8  & $ 57.2/91 $ \\
\hline
$L_3 L_2 E^c_3 $  & AMS-02 + Fermi &  3000 (fixed)  & 25.8 & $ 89.7/92 $ \\
\hline
\hline
$L_3 L_2 E^c_1 $(M1)  & AMS-02 + Fermi &  2276  & 26.3 & $ 65.7/91 $ \\
\hline
$L_3 L_2 E^c_2 $(M1)  & AMS-02 + Fermi &  4172  & 26.0 & $ 61.6/91 $ \\
\hline
$L_3 L_2 E^c_3 $(M1)  & AMS-02 + Fermi &  7887  & 25.8 & $ 56.8/91 $ \\
\hline

$L_3 L_2 E^c_1 $(M2)  & AMS-02 + Fermi &  2322  & 26.4 & $ 74.5/91 $ \\
\hline
$L_3 L_2 E^c_2 $(M2)  & AMS-02 + Fermi &  4979  & 26.1 & $ 72.1/91 $ \\
\hline
$L_3 L_2 E^c_3 $(M2)  & AMS-02 + Fermi &  5678  & 25.8 & $ 63.8/91 $ \\

\hline

\end{tabular}

\end{center}
\end{table}

For the results, we first notice that the decaying wino dark matter with $LLE^c$ interactions is indeed consistent with cosmic-ray electron and positron fluxes reported by AMS-02 and other collaborations. Furthermore, the most of the parameter space (the mass and width of the wino) explaining the positron anomaly is still consistent with gamma-ray observations, which is sharp contrast to the case of the decaying dark matter with $LH_u$ interactions. It is also clearly seen that the wino dark matter decaying into leptons of first and second generations requires the shorter lifetime than the one decaying into third generation leptons as expected. On the other hand, in some cases, the limits from the diffuse gamma-ray observations start excluding the parameter region. When the observation of its isotropic component is (officially) updated using the full data set that the Fermi-LAT collaboration already has, the decaying wino dark matter with some particular $LLE^c$ interactions such as $L_3L_2E_3^c$ will be favored or excluded. Even if the update result does not show the deviation from astrophysical backgrounds, the decaying wino dark matter into first and/or second generation leptons, such as the decay via $L_1 L_2 E^c_2$, will survive.

Another interesting thing we should point out is that the mass of the wino dark matter can be estimated from the results. It is trivial that the wino mass much lighter than 1~TeV is not consistent with the observed positron fraction. On the other hand, the wino mass much heavier than 1~TeV is also disfavored because the lifetime is too short to evade the gamma-ray constraints. Roughly speaking, the wino mass should be at around 3~TeV, which is nothing but the region the pure gravity mediation model predicts. Furthermore, with an appropriate $LLE^c$ interaction, the best fit value of the wino mass is just around 2.7~TeV, which is a very interesting coincidence because the pure wino dark matter predicts the mass of 2.7~TeV when its abundance observed today is from its thermal relic.

Finally, we would like to comment on a slight discrepancy between the result fitting to AMS-02 and that fitting to Fermi-LAT. The discrepancy is actually not due to our model but from the one between the two experimental results as already pointed out by recent studies~\cite{AMS-02vsFermi-LAT}. The essential reason is that the positron fraction that the AMS-02 collaboration reported is somewhat shallower than that reported by the PAMELA collaboration which is known to be consistent with the combined electron plus positron flux reported by the Fermi-LAT collaboration. As a result, for the decaying wino dark matter scenario, the wino lifetime to explain the AMS-02 result is always longer than that to explain the Fermi-LAT result, and it makes the discrepancy. The AMS-02 experiment will report on their new results of electron and positron fluxes in near future, and then we will have a clue to understand the discrepancy. Once we know the reason, more accurate analyses of the decaying wino dark matter will be possible.

\section{Summary}
\label{sec: summary}

In light of the recent result of the positron fraction reported by the AMS-02 collaboration, we have discussed the decay of the wino dark matter through $LLE^c$ interactions in the framework of the PGM model. The PGM model predicts the almost pure neutral wino as a lightest supersymmetric particle, namely a candidate for dark matter, whose mass is predicted to be around 1~TeV. This result is very interesting because the AMS-02 result in fact requires the dark matter mass around 1~TeV if the excess of the positron fraction is interpreted as a decaying dark matter signal. Furthermore, we have explicitly shown that the decaying wino through $LLE^c$ interactions well reproduce the AMS-02 result with being consistent with constraints from anti-proton and gamma-ray observations. We have also discussed a model of the R-parity violation where only leptonic R-parity violating interactions are generated in the appendix, while neither the other hadronic R-parity violating interactions nor bi-linear R-parity violating interactions are generated. Such a model is crucial to avoid the constraint from cosmic-ray anti-proton observations.

Here, we comment on some future prospects for testing the decaying wino dark matter. One of the distinct properties of the wino dark matter is that both its annihilation and decay provide signals to cosmic-ray and gamma-ray observations. This is sharp contrast to other dark matter candidates such as the decaying gravitino or stable particles whose decay are forbidden. The decaying wino dark matter can therefore be tested in various indirect searches. For instance, the wino produces anti-protons thorough its annihilation, which is expected to be detected at the AMS-02 experiment in near future~\cite{PGMs, spread}. As can be seen in previous section, the wino also produces diffused gamma-rays mainly through its decay, which can also be used to test the dark matter when data is accumulated enough.

Another interesting test for the decaying wino dark matter is the observations of galactic clusters and dwarf spheroidal galaxies using future gamma-ray telescopes. The decay process of the wino dominantly contributes to the signal from galactic clusters, which will be detected in the future at the next generation air-Cherenkov telescopes~\cite{Cirelli:2012ut}. On the other hand, the annihilation process of the wino becomes comparable to the decay process at the signal from dwarf spheroidal galaxies, which is also expected to be observed in future satellite experiments~\cite{G400}. When both signals are detected and their energy spectra are measured accurately, these signals from the above two processes have a potential to be a smoking-gun signature.

\vspace{1.0cm}

\noindent
{\bf Acknowledgments}
\vspace{0.1cm}\\
\noindent
This work is supported by the Grant-in-Aid for Scientific research from the Ministry of Education, Science, Sports, and Culture (MEXT), Japan (No. 24740151 for M.I., No. 23740169 for S.M. and No. 22244021 for S.M. \& T.T.Y), and also by the World Premier International Research Center Initiative (WPI Initiative), MEXT, Japan.

\appendix
\section{Possible origin of $LLE^c$ interactions}
\label{sec: LLE interactions}
In this paper, we have assumed that the R-parity violation is dominated by $LLE^c$ interactions. 
If we consider the GUT, however, $LLE^c$ interactions are accompanied by the other R-parity violating interactions $U^cD^cD^c$ and $QD^cL$ when they originate in the R-parity violating interaction $\mathbf{10} \, \mathbf{5^*} \, \mathbf{5^*}$. 
Here, we have used SU(5) GUT representations for the MSSM matter fields, $\mathbf{10} = (Q_L, \bar{U}_R, \bar{E}_R)$ and $\mathbf{5^*} = (\bar{D}_R,L_L)$. 
Note that those accompanied R-parity violating interactions do not cause rapid proton decay because of $\lambda = O(10^{-19})$ which is favored for the observed excess of the positron fraction.\footnote{This should be contrasted to the decaying gravitino dark matter scenario~\cite{Buchmuller:2007ui, Gravitino decay} which explains the excess of the positron fraction through the R-parity violating interactions. 
That is, since the decaying gravitino dark matter requires much larger R-parity violating $LLE^c$ interactions, the accompanied $U^cD^cD^c$ and $QD^cL$ interactions lead to a too short lifetime of a proton.} However, they cause a large flux of cosmic-ray anti-protons which easily exceeds the observed 
constraint on it~\cite{Garny:2012vt}. 

In this appendix, we discuss a model where $LLE^c$ interactions are accompanied by neither $U^cD^cD^c$ nor $QD^cL$ interactions in a GUT-consistent way. For this purpose, we consider an SU(5) $\times$ U(2)$_H$ product group GUT model~\cite{Product GUT}, where $LLE^c$ interactions originate not in 
$ \mathbf{10} \, \mathbf{5^*} \, \mathbf{5^*}$ but in R-parity violating interactions consisting of additional lepton-like fields, $L_H$, ${\bar L}_H$, $E_H$ and ${\bar 
E}_H$, in the U(2)$_H$ sector (see table 2).\footnote{For a related model of R-parity violation based on a product group GUT, see reference~\cite{Bhattacherjee:2013gr}.}

\begin{table}[t]
\begin{center}
\scriptsize
{\renewcommand\arraystretch{1.15}
\begin{tabular}{|c|c|c|c|c|c|c|c|c|c|c|c|}
\hline
& $\mathbf{10}$ & $\mathbf{5^*}$
& $\Phi(\mathbf{5})$ & $\bar{\Phi}(\mathbf{5}^*)$
& $X(\mathbf{1})$ & $\bar{X}(\mathbf{1})$
& $\Sigma({\mathbf 1})$
& $\bar{E}_H(\mathbf{1})$ & ${E}_H(\mathbf{1})$
& ${L}_{H}(\mathbf{1})$ & $\bar{L}_{H}(\mathbf{1})$
\\
\hline
SU(2)$_H$
& $\mathbf{1}$ & $\mathbf{1}$ 
& $\mathbf{2}$ & $\mathbf{2}$
& $\mathbf{2}$ & $\mathbf{2}$ 
& $\mathbf{3} + \mathbf{1}$
& $\mathbf{1}$ & $\mathbf{1}$
& $\mathbf{2}$ & $\mathbf{2}$
\\
\hline
U(1)$_H$
& $0$ & $0$ 
& $-1/2$ & $1/2$
& $-1/2$ & $1/2$ 
& $0$
& $1$ & $-1$
& $-1/2$ & $1/2$
\\
\hline
\end{tabular}
}
\end{center}
\caption{\small \sl Charge assignments on the SU(5) $\times$ U(2)$_H$ GUT gauge groups. 
Flavor indices of $\mathbf{10}$, $\mathbf{5^*}$, $L_H$, $\bar{L}_H$, $E_H$, and $\bar{E}_H$ are suppressed for simplicity. }
\label{tab: GUT}
\end{table}%

Let us first briefly review the product group GUT model based on SU(5) $\times$ U(2)$_H$. In this model, no adjoint of SU(5) is introduced and the GUT gauge symmetries are broken by the vacuum expectation values of bi-fundamental fields $\Phi$ and $\bar{\Phi}$ in table~\ref{tab: GUT}, $\langle \Phi_a^i \rangle = v \delta^i_a$ and $\langle \bar{\Phi}_i^a \rangle = v \delta^a_i$, where $v$ denotes a dimensionful parameter around the GUT scale and indices run as $a =$ 1--5 and $i =$ 1--2. Such vacuum expectation values are obtained as a supersymmetric solution of the superpotential:
\begin{eqnarray}
{\cal W} = \sqrt{2} \, \bar{\Phi}^a_j \, \Sigma^A \, (t^A)^j_i \, \Phi^i_a
+ \sqrt{2} \, \bar{X}_j \, \Sigma^A \, (t^A)^j_i \, X^i
- \sqrt{2} \, v^2 \, \Sigma^0,
\end{eqnarray}
where $t^A$ denotes Pauli matrices divided by two ($t^{1,2,3} = \sigma^{1,2,3}/2$) and the two-dimensional unit matrix divided by 2 ($t^0 = \mathbf{1}_{2 \times 2}/2$). With the vacuum expectation values, SU(5) $\times$ U(2)$_H$ are spontaneously broken down into MSSM gauge groups. 

As a remarkable feature of this model, MSSM Higgs doublets are not embedded in the fundamental representations of SU(5) but embedded in the doublets of U(2)$_H$, i.e. $H_d \sim \bar{\Phi}^a_i X^i$ and $H_u \sim \Phi_a^i \bar{X}_i$. Therefore, the doublet-triplet problem is absent since no colored Higgs multiplets are introduced in this model.\footnote{All the other modes consisting of $\Phi$'s and $\Sigma$ obtain masses at the GUT scale.} MSSM Yukawa interactions are effectively given by higher dimensional operators,
\begin{eqnarray}
{\cal W} = \frac{1}{M_*} \mathbf{10} \, \mathbf{5^*} \, \bar{\Phi} \, X
+ \frac{1}{M_*} \mathbf{10} \, \mathbf{10} \, {\Phi} \, \bar{X},
\end{eqnarray}
where flavor and gauge indices are suppressed. These higher dimensional operators may be generated by integrating out heavy particles, so that the dimensionful parameter $M_*$ is of the order of the GUT scale. 


Let us now introduce R-parity violation consisting of the hidden lepton-like fields,
\begin{eqnarray}
{\cal W} = \lambda_H L_{H}L_{H}\bar{E}_H ,
\end{eqnarray}
where couplings $\lambda_H$ is a dimensionless parameter. 
The hidden lepton-like fields have masses $M_L$ and $M_E$ which are assumed at 
GUT scale. 
Note that we need at least two pairs of the lepton-like doublets for non-vanishing R-parity violating interactions. 

After the GUT breaking, the R-parity violation of the lepton-like fields is mediated to the MSSM leptons through ${\cal W} = \kappa_{k} \,\Phi \, L_{k} \, \bar{L}_{H}$, 
which cause mixing between $L_H$  and $L$'s with the angles of $\kappa_{k} v / M_L$. In addition, we also consider higher dimensional operators, ${\cal W} = (\kappa'_k/M_*) \, \mathbf{10}_k \, \bar{\Phi} \, \bar{\Phi} \, E_H = (\kappa'_k v^2/M_*) E^c_k \, E_H$, which lead to 
the mixing angles between $E_H$ and $E$'s, $\kappa'_i v^2/(M_* M_E)$. Through these mixings, R-parity violating interactions of the lepton-like fields generate the required leptonic R-parity violating operator,
\begin{eqnarray}
{\cal W} = 
\frac{{\lambda}_H \, \kappa_i \, \kappa_j \, \kappa'_k \, v^4}
{M_L^2 \, M_E \, M_*} L_i L_j E^c_{k},
\end{eqnarray}
while $U^cD^cD^c$ and $QD^cL$ interactions are suppressed. In this way, we obtain a model in which only the leptonic R-parity violation is realized in a GUT consistent manner.
For $\kappa_k ={\cal O}(1)$, $M_L \sim M_E \sim v \simeq 10^{16}$\,GeV and $M_* \simeq 10^{17}$\,GeV,
and $\lambda_H \simeq 10^{-18}$, we obtain the required R-parity violation $\lambda\simeq 10^{-19}$.%
\footnote{
One may consider a simpler model where no $E_H$'s are introduced.
In this case, the $LLE^c$ originates in an R-parity violating interaction,
$\mathbf {10} \bar{\Phi }L_H
\bar{\Phi}L$ through the mixing between $L$ and $L_H$ 
without accompanied by the problematic $U^cD^cD^c$ and $QD^cL$ interactions.
In this simpler model, one may  suppress other R-parity violating terms by assuming
$Z_{10R}$ symmetry with charge assignments, 
${\mathbf{10}}(2)$, 
${\mathbf{5}^*}(-2)$, 
${\Phi}(-2)$,
${\bar{\Phi}}(2)$,
$\Sigma (2)$,
$X(0)$, $\bar{X}(0)$, 
$L_H(-2)$, 
$\bar{L}_H(2)$,
and the right-handed neutrino $N_R^c(6)$,
which allows the see-saw mechanism~\cite{seesaw}.
}


\end{document}